\documentclass[aip,amsmath,amssymb
,reprint,sort&compress]{revtex4-1}

\usepackage{graphicx}
\usepackage{dcolumn}
\usepackage{bm}
\usepackage[utf8]{inputenc}
\usepackage[T1]{fontenc}
\usepackage{mathptmx}
\usepackage{lineno,hyperref}
\usepackage{subcaption}
\usepackage{siunitx}
\DeclareSIUnit\year{a}
\usepackage{placeins}
\modulolinenumbers[5]
\usepackage{caption}
\begin{document}

\preprint{AIP/123-QED}

\title[Optimization of a laser ion source for $^{163}$Ho isotope separation]{Optimization of a laser ion source for $^{163}$Ho isotope separation}

\author{Tom Kieck}
\affiliation{Institute of Nuclear Chemistry, Johannes Gutenberg University, 55128 Mainz, Germany}
\affiliation{Institute of Physics, Johannes Gutenberg University, 55128 Mainz, Germany}
\email{tomkieck@uni-mainz.de}

\author{Sebastian Biebricher}
\affiliation{Institute of Physics, Johannes Gutenberg University, 55128 Mainz, Germany}

\author{Christoph E. D\"{u}llmann}
\affiliation{Institute of Nuclear Chemistry, Johannes Gutenberg University, 55128 Mainz, Germany}
\affiliation{GSI Helmholtzzentrum f\"{u}r Schwerionenforschung GmbH, 6929 Darmstadt, Germany}
\affiliation{Helmholtz-Institut Mainz, 55099 Mainz, Germany}

\author{Klaus Wendt}
\affiliation{Institute of Physics, Johannes Gutenberg University, 55128 Mainz, Germany}

\date{\today}

\begin{abstract}
To measure the mass of the electron neutrino, the "Electron Capture in Holmium-163" (ECHo) collaboration aims at calorimetrically measuring the spectrum following electron capture in $^{163}$Ho. The success of the ECHo experiment depends critically on the radiochemical purity of the $^{163}$Ho sample, which is ion-implanted into the calorimeters. For this, a \SI{30}{\kilo\volt} high transmission magnetic mass separator equipped with a resonance ionization laser ion source is used. To meet the ECHo requirements, the ion source unit was optimized with respect to its thermal characteristics and material composition by means of finite element method (FEM) thermal-electric calculations and chemical equilibrium simulation using the Gibbs energy minimization method. The new setup provides a much improved selectivity in laser ionization versus interfering surface ionization of \num{2700(500)} and a superior overall efficiency of \SI{41(5)}{\percent} for the ion-implantation process.

\end{abstract}

\maketitle


\section{Introduction}
\begin{figure*}
	\includegraphics[width=\textwidth]{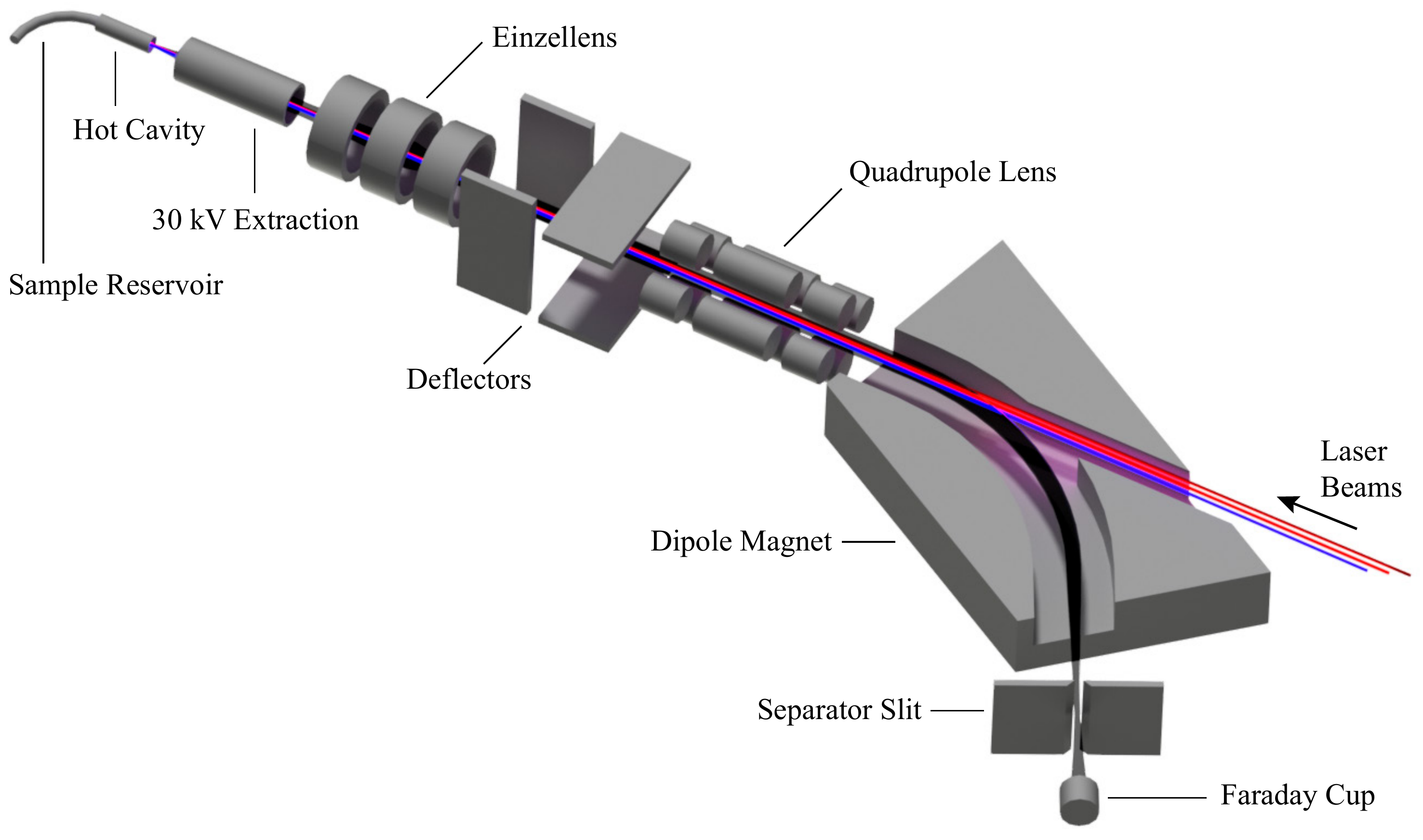}
	\caption{Overview of the RISIKO mass separator.}
	\label{fig:risiko}
\end{figure*}

A resonance ionization laser ion source (RILIS) provides an element-selective and remarkably efficient way to produce ions of almost any element. Such sources are widely used at isotope separator on-line (ISOL) facilities for the efficient production of isotopically and isobarically pure ion beams of rare isotope \cite{Letokhov1979,Koster2003,Fedosseev2017}. Already early it has been shown that multi-step laser excitation is particularly useful for selective and efficient isotope separation in the rare earth elements \cite{Karlov1978}. Therefore, this process is also well suited for ion production to separate and ion-implant the rare unstable isotopes $^{163}$Ho \cite{Schneider2016} used in the ECHo \cite{Gastaldo2017} (Electron Capture in $^{163}$Ho) project. This aims at measuring the electron neutrino mass by recording the spectrum following electron capture of $^{163}$Ho using metallic magnetic calorimeters (MMCs) \cite{Gastaldo2013}. The synthetic radioisotope $^{163}$Ho has a half-life of \SI{4570(50)}{\year} \cite{Baisden1983}. For ECHo, it is produced artificially by intense neutron irradiation of enriched $^{162}$Er samples in the ILL high flux nuclear reactor. The $^{163}$Ho is then chemically separated from the target material and unwanted byproducts \cite{Dorrer2018}. The obtained sample mainly consists of the Ho isotopes 163, 165, and 166m. While $^{165}$Ho is stable, the latter isomer $^{166m}$Ho is a long-lived ($T_{1/2} = \SI{1132(4)}{\year}$ \cite{Nedjadi2012}) radioisotope. If present in the MMCs, its emitted radiation will lead to undesired background in the spectrum of $^{163}$Ho. Therefore, the sample is mass separated at the RISIKO isotope separator at Johannes Gutenberg University Mainz to ensure a single-isotope ion implantation into the Au-absorber arrays of the ECHo MMCs. The RISIKO facility in its application for the ECHo source implantation is described in \cite{Schneider2016} including all details on the laser system and the all-resonant three-step laser excitation and ionization process, which is carried out along strong optical resonance lines of the Ho atom. A sketch of the major components of the mass separator is provided in figure \ref{fig:risiko}.

For future phases of the ECHo experiment large amounts of $^{163}$Ho of up to \num{1e17} atoms are intended to deliver suitable statistics in the decay spectrum \cite{Gastaldo2017}. Correspondingly, for the ion implantation high ion currents in the order of \SI{1}{\micro\ampere} and beyond for long term operation over months will be needed. Because of the limited availability of $^{163}$Ho, an excellent efficiency of the ion source as well as the overall implantation process is required. For this purpose the RISIKO mass separator is equipped with an advanced RILIS, allowing for dedicated and well controlled insertion, evaporation, atomization, and final ionization of a multitude of elements including Ho under appropriate conditions. High mobility and negligible losses of the often precious and irrecoverable sample material which is often available only in limited quantities, inside the ion source unit during evaporation and transfer from the sample reservoir to the location of ionization by the entering laser radiation is essential. In addition, highest reliability, mechanical stability during long term operation and re-usability of the major ion source components are indispensable. All these features depend on the physical, thermal, and chemical properties of the sample material, the carrier foil (which is introduced in most cases to assist complete chemical reduction of the sample), and the mechanical ion source components that were studied an optimized in this work. Different surface ion sources for on-line applications were already analyzed and optimized with similar methods \cite{Alton2003,Manzolaro2010,Raeder2014,Manzolaro2016} but results from these sources are not directly transferable to the RISIKO RILIS. At on-line facilities the massive proton target for production of exotic nuclides by bombardment with highly energetic projectiles leads to a significantly differing thermal situation regarding heat dissipation and distribution in comparison to the reservoir/cavity arrangement of the RISIKO laser ion source setup for the ECHo implantations. Substantial molecule formation in on-line sources is caused by the target material (e.g. UC$_{x}$) and cannot be suppressed sufficiently by adding reducing agents as in case of the off-line RILIS at RISIKO.

In general, multi-step resonant laser ionization is entirely element selective but influenced by a number of disturbances and unwanted side effects. Ionization on the hot surfaces \cite{Kirchner1990} that are necessary to produce and sustain single-atom metal vapors contribute to ion-formation aside of resonance ionization. The influence of this effect depends on the purity of the initial sample and on the carrier foil material. In addition, upon heating the sample, evaporation may partially occur in the form of molecular species. They may either decompose, thus releasing atomic species accessible for resonance ionization, or ionize on hot surfaces as molecules. At present, different techniques to overcome this problem of unwanted contaminations in the ion beam have been developed. These include decoupling the ionization volume from the atomizer (e.g. in the LIST or IG-LIS approach \cite{Fink2013,Raeder2014}) or using dedicated low-work function materials for lining the inside of the ion source \cite{Schwellnus2009}. Main arguments against using these techniques in case of the ECHo ion implantation are high sample losses of typically around \SI{90}{\percent} and low ion beam currents due to temperature limitations with rapid degradation of the ion source material in use.

\section{Thermal-Electric Simulation}
\subsection{Model and Implementation}
\begin{figure}
	\centering
	\includegraphics[width=0.4\textwidth]{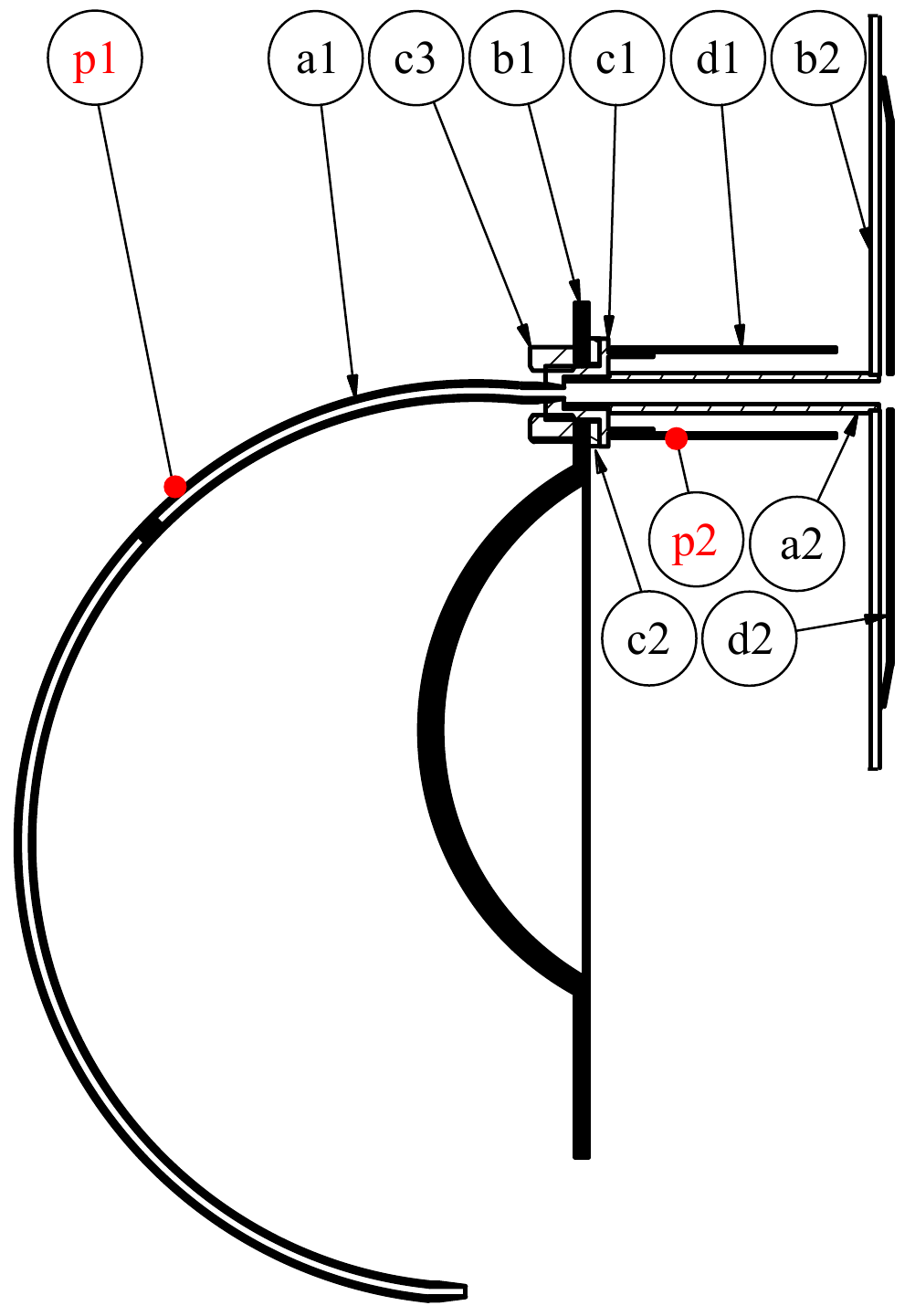}
	\caption{Cross section through the RISIKO laser ion source version 1. The parts are described in the text. Points of pyrometric measurements for chapter \ref{chapter:measurement} are indicated as p1 and p2.}
	\label{fig:ion-source-cut}
\end{figure}

The conventional RISIKO laser ion source (denoted here as version 1 and sketched in figure \ref{fig:ion-source-cut}) which was the starting point for the current work, consists of two main parts that are resistively heated independently by electric direct currents. Part 1, the sample reservoir (a1) serves to produce, transport, and confine the sample vapor and is jointly connected to part 2, the hot cavity (a2), wherein the atoms are ionized. The sample is placed in the upper part of the reservoir, a \SI{165}{\milli\meter} long tantalum capillary with \SI{1}{\milli\meter} inner diameter, closed in the upper third. In the adjacent hot cavity (a2), a \SI{35}{\milli\meter} tantalum tube with \SI{2,5}{\milli\meter} inner diameter, the lasers are focused to resonantly excite and ionize the atoms. Generated ions are guided by the electric field of the heating currents towards the exit hole to access the electrostatic extraction field of the mass separator. Heating currents (max. \SIlist{150;400}{\ampere} for reservoir and hot cavity, respectively) are applied via an air-cooled clamp to the end of the reservoir and from a water-cooled rod through a multilayer tantalum spring (b1) to the hot cavity. Both are grounded through the electrode (b2), which is mounted on a water-cooled plate. The tantalum mounting part (c1) is attached to the spring using a washer (c2) and nut (c3), both made of graphite, to prevent hot surface welding. Heat radiation of the hot cavity is shielded radially by a tantalum tube (d1) surrounded by a multilayer foil and an additional tantalum sheet (d2) towards the extraction gap.

\begin{figure}
	\centering
  \begin{subfigure}[b]{0.225\textwidth}
        \includegraphics[width=\textwidth]{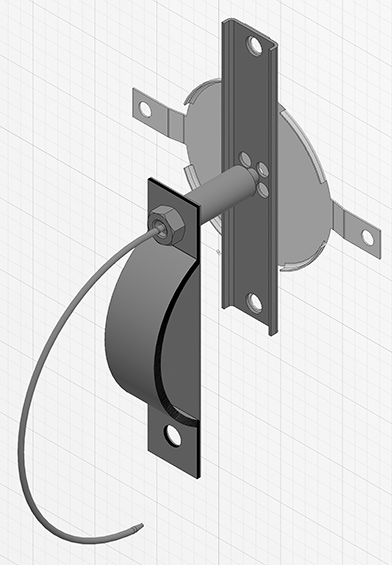}
				\caption{before preprocessing}
  \end{subfigure}
  ~ 
  \begin{subfigure}[b]{0.225\textwidth}
      \includegraphics[width=\textwidth]{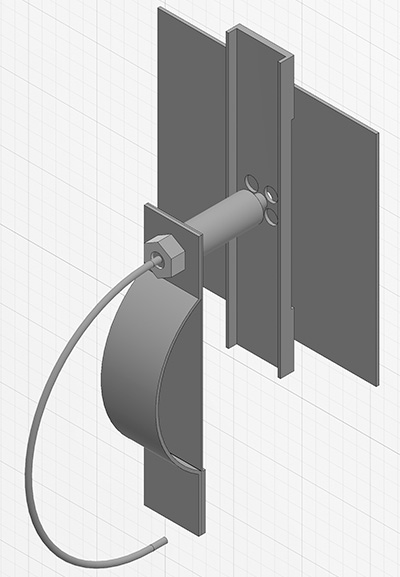}
			\caption{after preprocessing}
  \end{subfigure}
  \caption{3D CAD model of the RISIKO laser ion source version 1.}
	\label{fig:cad-comparison}
\end{figure}

The ion source is drawn as a 3D CAD model in detail. Before computing the temperature distribution a preprocessing step was performed: the model was stripped of irrelevant components and shapes to lower the geometric complexity and save computer run-time. Especially round shapes not significantly affecting the power dissipation or heat transfer were transformed to preferably parallel-edged shapes leading to lower node densities in meshing the finite elements. Figure \ref{fig:cad-comparison} shows the ion source version 1 before and after preprocessing.

The finite element method calculations are carried out by Autodesk\textsuperscript{\textregistered} CFD \footnote{Autodesk\textsuperscript{\textregistered} CFD 2019 Build 20180307} with temperature-dependent material properties \cite{Hixson1992,Milosevic1999,Neuer1995,Rasor1960,Taylor1969}. Three thermodynamical effects are considered to obtain reliable results \cite{DieterBaehrKarlStephan2009}:

\begin{enumerate}
	\item The heating power $P$ which is produced by a constant current $I$, assuming a resistivity $R$ depending on the temperature $T$ as
		\begin{align*}
			P(T) = R(T) \cdot I^2.
		\end{align*}
	\item The transfer of heat $q$ by thermal conduction given by
		\begin{align*}
			\dot{q} = -\lambda(T) \nabla T,
		\end{align*}
		where $\lambda$ is the temperature-dependent thermal conductivity.
	\item The heat radiation acting as heat transport
		\begin{align*}
			\dot{q} = \varepsilon(T) \sigma T^4,
		\end{align*}
		with temperature-dependent hemispherical emissivity $\varepsilon$ and the Stefan-Boltzmann constant $\sigma$.
\end{enumerate}

\begin{figure}
	\centering
	\includegraphics[width=0.45\textwidth]{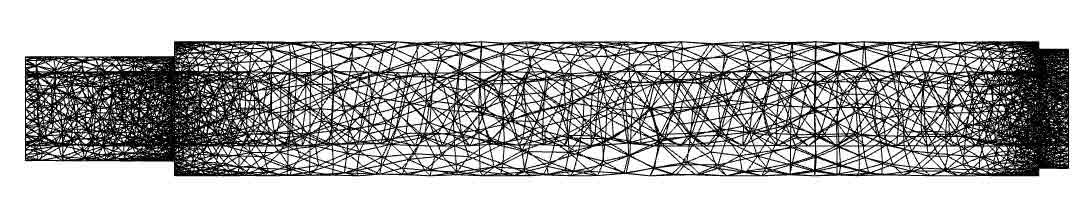}
	\caption{Adaptive mesh of the hot cavity (a2) with higher node density in the regions of high temperature gradient because of contact to cooler adjacent parts.}
	\label{fig:mesh-adaptive}
\end{figure}

Numerical approximations of the differential equations used in the simulation need to be analyzed for convergence in node distance $\Delta x = |x_A - x_B|$ of the mesh and iteration steps $n$ \cite{Hutton2004}. For the first calculation step a rough standard mesh is created and the solver runs in steady-state mode at highest heating currents until convergence is reached. In three subsequent steps, areas of high temperature gradients are localized and the mesh in these regions is refined to enhance accuracy. The resulting adaptive mesh, as shown exemplarily for the hot cavity in figure \ref{fig:mesh-adaptive} with about \num{500000} elements is used for the subsequent calculations, in which the heating currents are varied.

\subsection{Simulation Results}

\begin{figure}
	\includegraphics[width=0.45\textwidth]{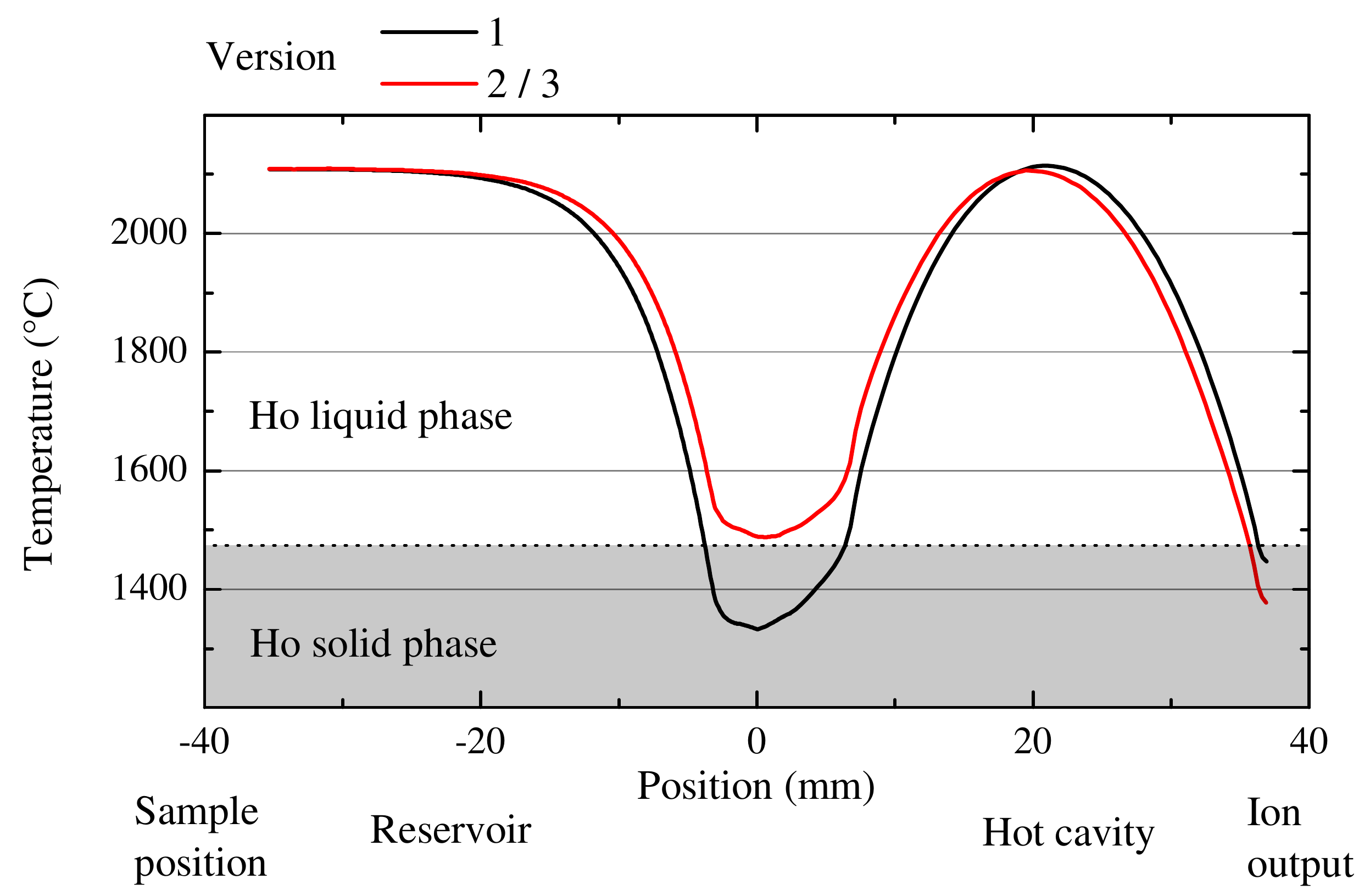}
	\caption{Temperature curve of the discussed ion source versions. The heating power for reservoir and hot cavity is \SIlist{590;340}{\watt} for version 1, \SIlist{590;325}{\watt} for version 2, and \SIlist{290;325}{\watt} for version 3. The melting point of Ho (\SI{1472}{\degreeCelsius} \cite{Spedding1960}) is indicated by the dashed line.}
	\label{fig:temperature-comparison}
\end{figure}

The result of the initial design version 1 shows a significant temperature drop in the connection region between sample reservoir and hot cavity, where significant heat losses occur through the tantalum spring and adjacent parts. The overall temperature distribution along the source unit is shown in figure \ref{fig:temperature-comparison} by the black curve. This leads either to a condensation point for the sample vapor at this position or to a high surface ion background and mechanical instability while overheating the whole setup to raise the coldest spot above the sample melting point. One aspect of primary importance for optimum source performance was thus to modify the design in such a way that the coldest temperature in this region could be kept above the melting point of Ho of \SI{1472}{\degreeCelsius} \cite{Spedding1960}.

Assuming highly efficient laser ionization, the region near the ion source exit hole is only of very minor relevance for possible condensation effects. The atoms are already ionized at the reservoir exit and ions are confined by a negative thermal plasma potential \cite{Liu2011}. Correspondingly they are efficiently transported to the extraction gap by the weak electric field of the properly poled heating potential along the hot cavity. The refined version 2 of the RILIS evidently includes three improvements. At the central current support between reservoir and hot cavity (c1) to significantly reduce heat-loss through thermal radiation:
\begin{enumerate}
	\item The spring (b1) is cut off at the top to the smallest possible geometry conserving current and mechanical stability.
	\item Because of the very high emissivity of graphite ($\epsilon \approx 0.85$ \cite{Neuer1995}), the material of nut (c3) is changed to molybdenum.
	\item The graphite washer (c2) is also replaced by one made of molybdenum.
\end{enumerate}

The resulting temperature curve for version 2 is rather similar to before but now every point of the atom path from reservoir to hot cavity has a temperature above the melting point of Ho as indicated in figure \ref{fig:temperature-comparison} by the red line.

\begin{figure}
	\centering
	\includegraphics[width=0.45\textwidth]{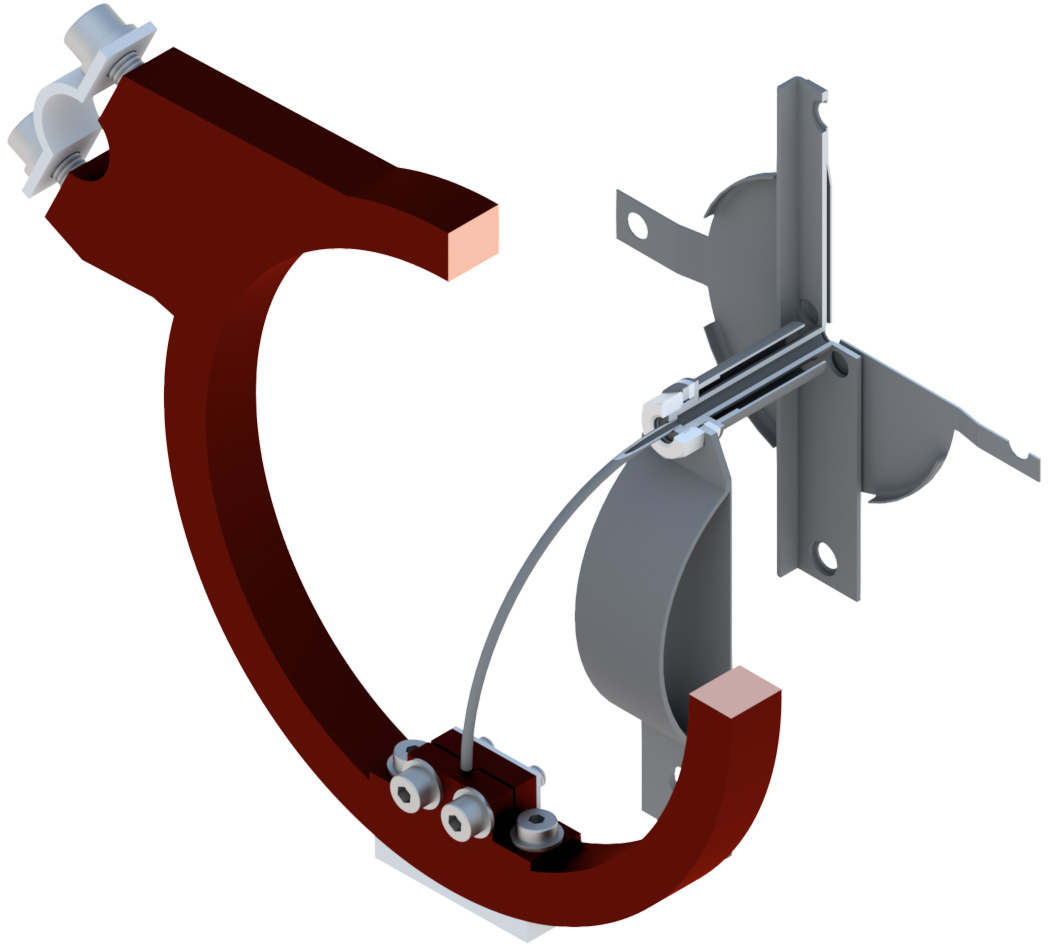}
	\caption{Computer-rendered drawing of the ion source version 3 with a 1/4 cutout to see the inner structure.}
	\label{fig:ion_source_3}
\end{figure}

In the next optimization step towards version 3, special attention was given to the mechanical stability of the reservoir (a1) to prevent any possible sample-material losses or measurement disruptions by distortion or rupture of ion source components. A copper supporting ring was constructed that uses the existing infrastructure of two water cooled copper rods for a stable positioning and optimum cooling of the rear end of the reservoir as shown in figure \ref{fig:ion_source_3}. The sample reservoir was shortened to half its original length and now consists of a \SI{90}{\degree} bent capillary, held securely on the copper supporting ring. This design allows for a reliable connection of the reservoir to the hot cavity also at high temperatures, where Ta is getting soft. It also significantly reduces the heat generation at the ion source, while fully conserving the thermal profile of version 2 along the atom path. The related heat stress of the various materials inside the vacuum chamber is thus minimized.

\section{Chemical Equilibrium Simulation}

As second step, the high-temperature chemistry of the different interacting materials inside the ion source unit was studied. The chemical composition of the sample material is of highest relevance for the achievable RILIS efficiency. Numerous studies have shown that optimum results are obtained, if small samples are dissolved in diluted nitric acid for introduction into the source. Therein the majority of elements, predominantly the metals, are present as oxides (e.g. Ho$_2$O$_3$). Before insertion into the ion source, the sample is dried on a small piece of carrier foil of typically \SI{3 x 3}{\milli\meter} size, which is co-introduced to serve as reducing agent. The overall mass separator ion current emitted from the ion source is composed of the total of all ionized species. Correspondingly, the ion beam quality in respect to emittance and sample throughput can be maximized by minimizing the ionization of unwanted contaminations on hot surfaces of the ion source. Accordingly, at high temperatures around \SI{2100}{\degreeCelsius} and low pressure of about \SI{e-6}{\milli\bar}, beside the reduction capability of the carrier foil also its high temperature characteristics has to be considered. Optimum properties include a low vapor pressure and a high ionization potential.

For minimizing the Gibbs energy of the chemical reaction system between sample and reducing agent \cite{White1958} the Equilibrium Composition Module of Outotec HSC Chemistry \footnote{Outotec HSC Chemistry 9.4.1 Build 09.01.2018} is used. The results are interpreted considering that this type of chemical simulation does not take into account reaction kinetics. Obtained results indicate that binary compounds of lanthanide with nitrogen are energetically favorable. Experimentally they were never observed and seem to be strongly inhibited. Thus, these compounds were excluded from the simulation.

\begin{figure}
	\centering
	\includegraphics[width=0.45\textwidth]{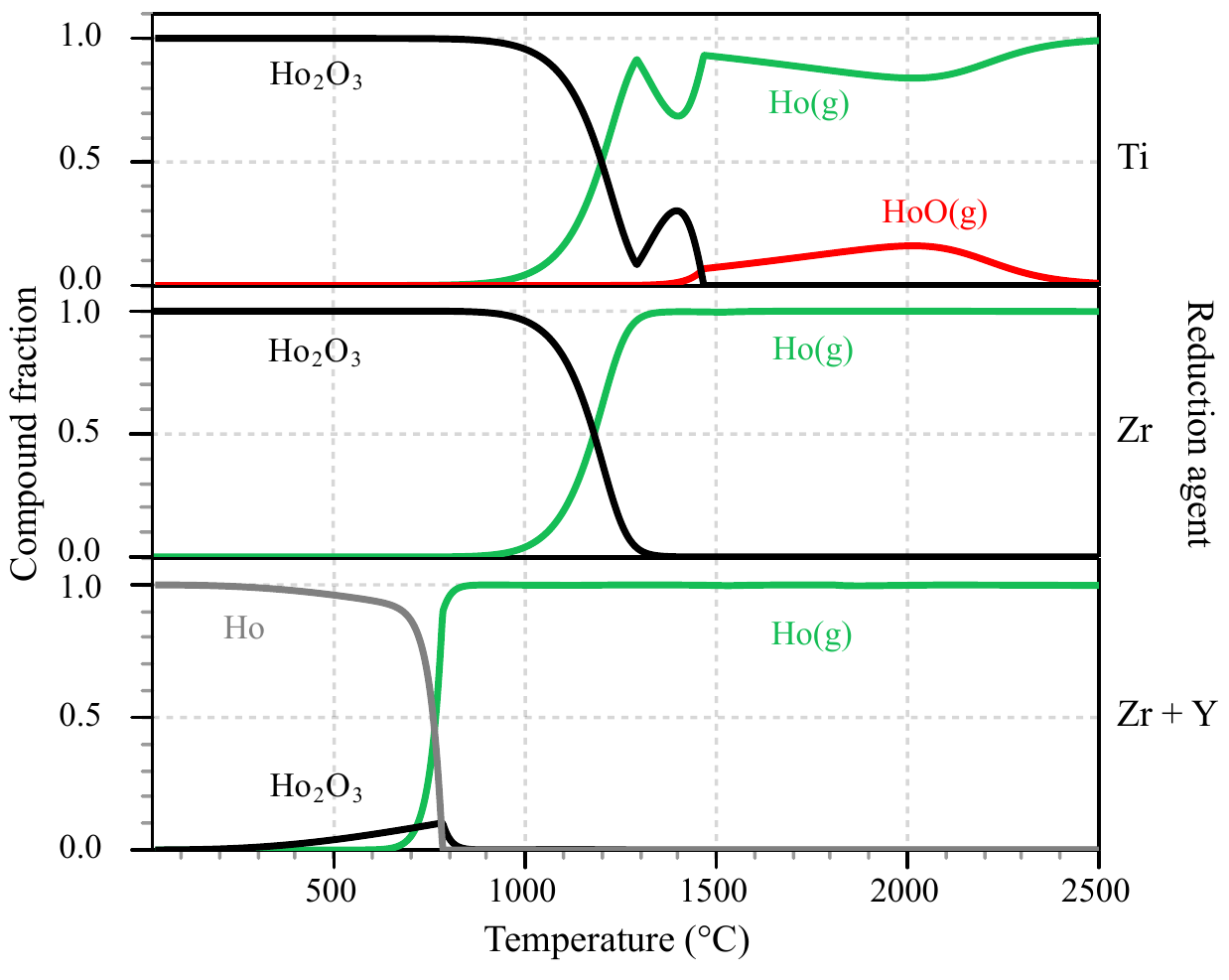}
	\caption{Temperature dependent formation of Ho compounds with different reducing agents, obtained with Outotec HSC Chemistry.}
	\label{fig:reduction}
\end{figure}

The standard carrier foil material for the RISIKO laser ion sources, i.e. Ti of \SI{12,5}{\micro\meter} thickness, shows disadvantages in the temperature regime needed for high ion currents starting at around \SI{1500}{\degreeCelsius} \cite{Schneider2016} by significant HoO gas formation as shown in figure \ref{fig:reduction}. The molecules are either not ionized or lost in the magnetic separation. This effect can be avoided by using Zr, which has the additional advantage of a higher melting point. The addition of a small amount of Y in the same order of magnitude as the Ho sample material lowers the reduction temperature and leads to a substantially more uniform and slightly slower rise of the ion current during ion source heating up and thus contributes to a better control of the ion source output and a reduction in high temperature interferences.

\section{Experimental Results}

The different adaptations from the simulation results were extensively tested before implementation for routine use in the different implantation phases of the ECHo project and for other experiments at the RISIKO mass separator. For this purpose additional measuring capabilities regarding various relevant parameters and their systematics were introduced.

\subsection{\label{chapter:measurement}Temperature Measurement}

A two color pyrometer (Heitronics KT 81R) is used for a contactless temperature measurement under full operation conditions . While measuring at two infrared wavelengths the results are widely independent of the vacuum window properties and any kind of pollution because of source material vaporization and condensation. The material-dependent emissivity ratio was precisely calibrated by measuring temperatures of a Ta wire up to the melting point with an \SI{10}{\degreeCelsius} stat. and \SI{22}{\degreeCelsius} sys. error. 

\begin{figure}
	\centering
	\includegraphics[width=0.45\textwidth]{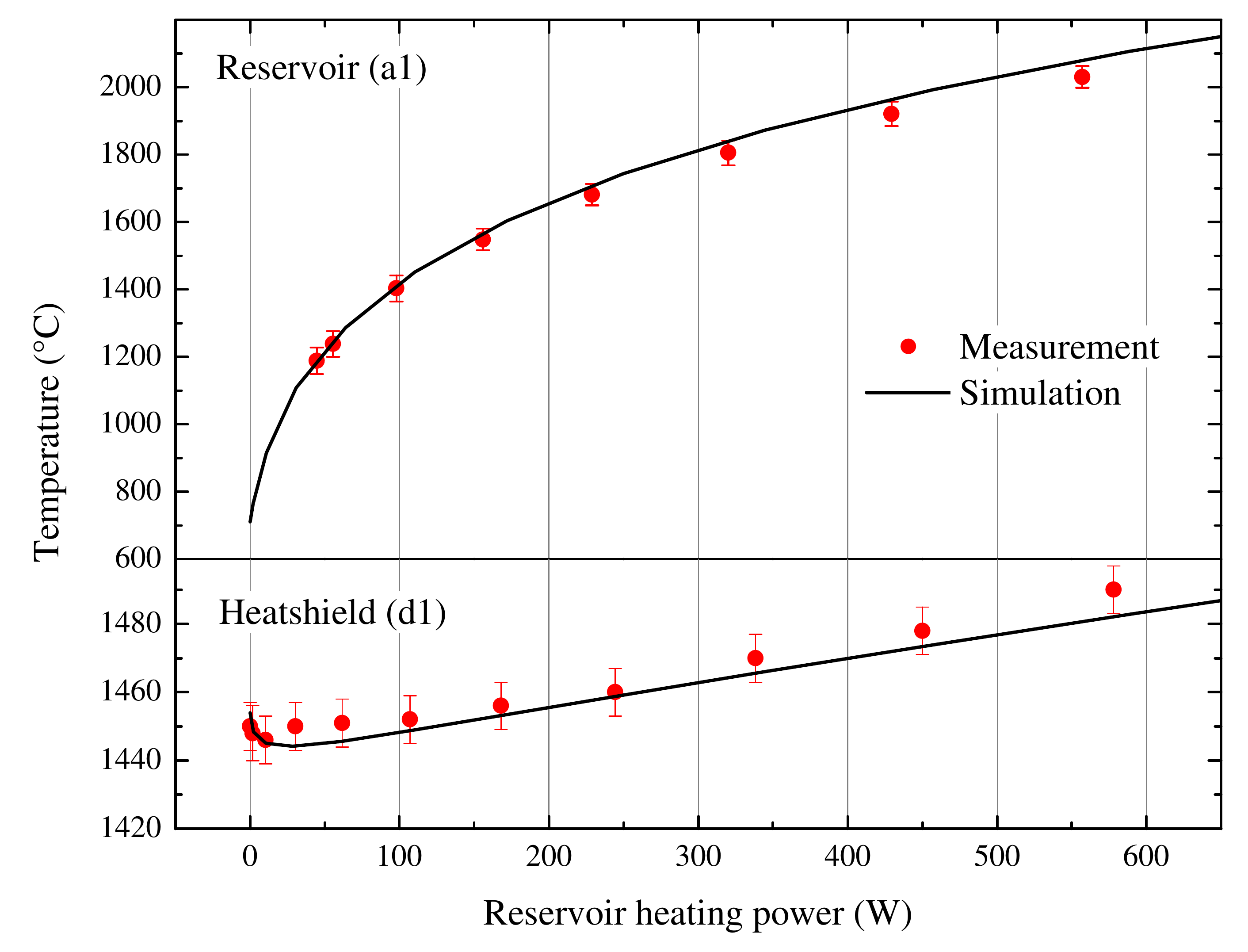}
	\caption{Comparison of ion source version 2 temperature measurement and simulation result at the sample reservoir and the hot cavity heat shield. The hot cavity is heated with \SI{325}{\watt}.}
	\label{fig:temp-meas}
\end{figure}

The measurement was performed at two characteristic positions of the ion source version 2: heat shield (p1) and sample position in the reservoir (p2) as indicated in figure \ref{fig:ion-source-cut}, combined with heating-power measurements of reservoir (a1) and hot cavity (a2) inside the vacuum chamber. The comparison of measurement and simulation at the reservoir is given in figure \ref{fig:temp-meas}, showing a very good absolute temperature agreement within uncertainties. To compare the relative effects of heat radiation and conduction, a measurement of the heat shield is given too, which also agree with results from the simulation.

\subsection{Ionization Efficiency}

\begin{figure}
	\centering
	\includegraphics[width=0.45\textwidth]{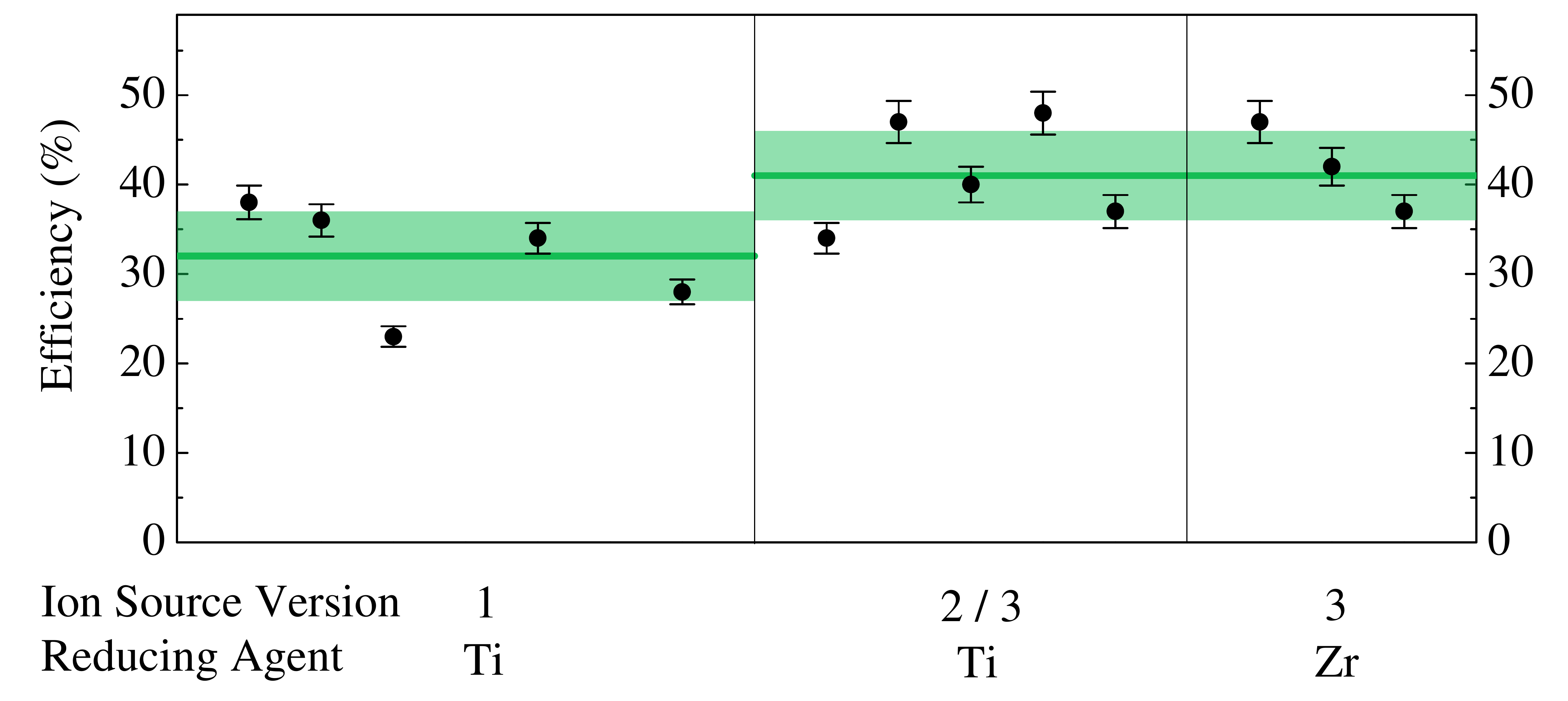}
	\caption{Overview of the ionization efficiency measurements. The average efficiencies of \SIlist{32(5);41(5)}{\percent} with standard deviation are indicated by the shaded area.}
	\label{fig:efficiency}
\end{figure}

To evaluate the effect of temperature profile optimization and homogenization, overall efficiencies as described in \cite{Schneider2016} with the different ion source versions 1 to 3 were measured and compared in figure \ref{fig:efficiency}. The \SI{32(5)}{\percent} mean value of the earlier measurements with version 1 is increased to \SI{41(5)}{\percent} of version 2 and 3. This is in good agreement with a \SI{40}{\percent} value, as obtained independently by Liu with a slightly different ion source type \cite{Liu2014}. This value seams to be close to the efficiency limit of the particular laser ionization scheme for Ho, no longer being influenced by the ion source geometry anymore. The newly chosen reducing agent of Zr or even Zr + Y does not show any significant influence to the overall efficiency and thus gives no hint towards a reduced HoO gas formation.

\subsection{Mass Spectrometry}

\begin{figure}
	\centering
	\includegraphics[width=0.45\textwidth]{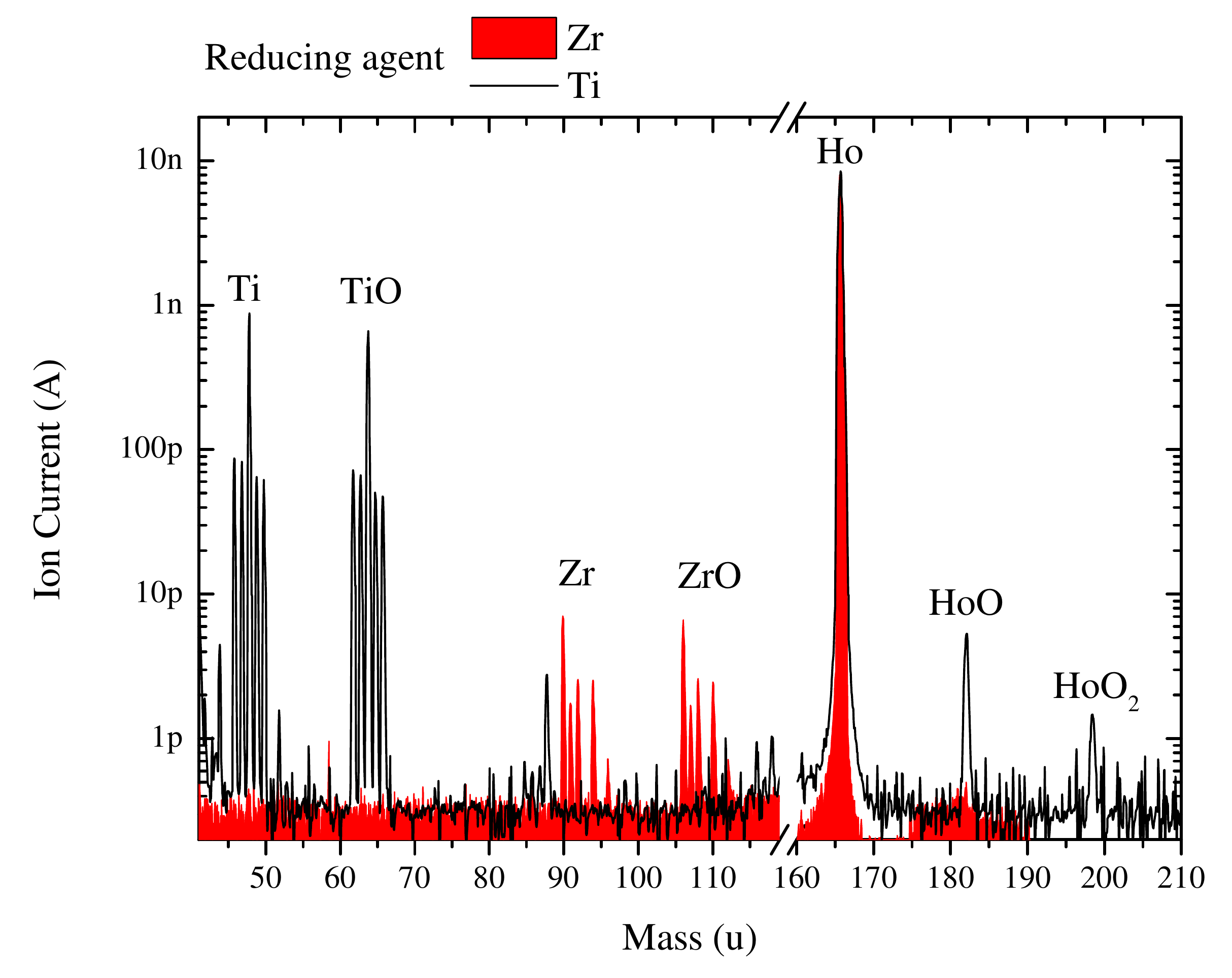}
	\caption{Relevant parts of the mass spectrum by resonance ionization of Ho with residual surface ionization, measured at reservoir temperatures of about \SIlist{1550}{\degreeCelsius}.}
	\label{fig:massscan}
\end{figure}

Surface ionization gives the possibility to compare the chemical simulation results with the experimental situation. Depending on the ionization potential almost every species in the hot cavity is ionized to a certain extent. Figure \ref{fig:massscan} compares the measured mass spectra for resonance ionization of Ho with residual surface ionization for the two reducing agents Ti and Zr. The mass spectrum with Ti shows peaks of Ho, HoO, HoO$_2$ and Ti, also with its oxide. Due to the low melting point of Ti, the number of all Ti ions is almost in the same order of magnitude as that of Ho ions. This might affect the mass separator performance and transmission at high ion currents. The mass spectrum with Zr as reducing agent shows a negligible amount of ions beside Ho. The absence of HoO confirms the excellent reduction capability of Zr as predicted by the simulation.

\subsection{Laser Selectivity}

\begin{figure}
	\centering
	\includegraphics[width=0.45\textwidth]{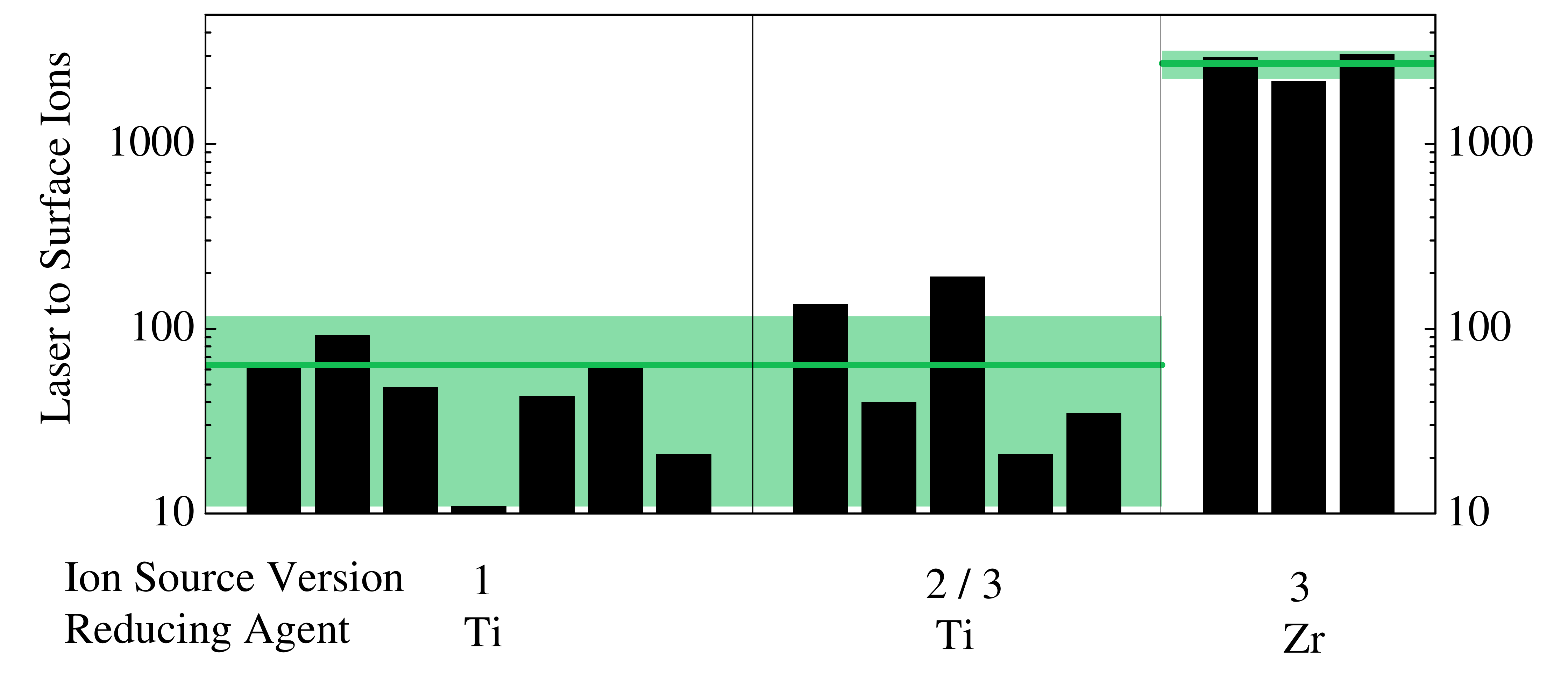}
	\caption{Overview of the laser-to-surface ionization selectivity measurements. The average ratios of \numlist{60(50);2700(500)} with standard deviation are indicated by the shaded area.}
	\label{fig:selectivity}
\end{figure}

A further way to analyze the chemical composition inside the ion source is to measure the surface-ion current while blocking the laser beams and compare it to the laser ion signal. These measurements are shown in Figure \ref{fig:selectivity}. A fraction of the HoO gas molecules that hits the hot cavity walls is reduced and surface ionized independently of the laser light. This effect leads to a rather low laser-to-surface ion ratio of \num{60(50)} (see figure \ref{fig:selectivity}). This signal is most significant at high ion current as produced at temperatures of about \SI{1500}{\degreeCelsius} confirming the chemical simulation in terms of HoO gas formation. Zr as reducing agent significantly lowers the surface ion signal and leads to a laser ionization selectivity of \num{2700(500)} in the ratio of laser to surface ions, almost 50 times surpassing the earlier value.

\section{Conclusion}
Extensive thermal and chemical simulations have been performed to improve the off-line RILIS used at the RISIKO mass separator of Johannes Gutenberg University Mainz for the present and upgrading it for the future phases of the ECHo experiment. Temperature and mass spectrometric measurements confirm the simulation predictions and the usefulness of the corresponding experimental improvements. The final ion source design has shown highest reliability in operation and an excellent efficiency in Ho ionization of \SI{41(5)}{\percent}, in good agreement with the value \SI{40}{\percent} of Liu obtained at the ISTF2 facility of ORNL in a rather similar approach \cite{Liu2014}. The newly chosen reducing agent Zr increases the laser-to-surface ionization selectivity by almost two orders of magnitude, and thus prevents significantly disturbances from interfering surface ions even at high ion source temperatures as needed for envisaged high ion currents in the \si{\micro\ampere} range.

\section{Acknowledgments}
We thank Yuan Liu from ORNL for the fruitful discussions about ion source simulations. This work was performed in the framework of DFG Research Unit FOR 2202 (ECHo), and we gratefully acknowledge the financial support under contract DU 1334/1-1.

\FloatBarrier

\section*{References}

\bibliography{mybibfile}

\end{document}